# Estimating the Accuracy of the Return on Investment (ROI) Performance Evaluations

**Alexei Botchkarev**
GS Research & Consulting and
Adjunct Prof., Ryerson University, Toronto, Canada
alex.bot@gsrc.ca

## Abstract

Return on Investment (ROI) is one of the most popular performance measurement and evaluation metrics. ROI analysis (when applied correctly) is a powerful tool in comparing solutions and making informed decisions on the acquisitions of information systems. The purpose of this study is to provide a systematic research of the accuracy of the ROI evaluations in the context of the information systems implementations. Measurements theory and error analysis, specifically, propagation of uncertainties methods were used to derive analytical expressions for ROI errors. Monte Carlo simulation methodology was used to design and deliver a quantitative experiment to model costs and returns estimating errors and calculate ROI accuracies. Spreadsheet simulation (Microsoft Excel spreadsheets enhanced with Visual Basic for Applications) was used to implement Monte Carlo simulations. The main contribution of the study is that this is the first systematic effort to evaluate ROI accuracy. Analytical expressions have been derived for estimating errors of the ROI evaluations. Results of the Monte Carlo simulation will help practitioners in making informed decisions based on explicitly stated factors influencing the ROI uncertainties.

# 1.0 Introduction

Return on Investment (ROI) is one of the most popular performance measurement and evaluation metrics. ROI analysis (when applied correctly) is a powerful tool in making informed decisions on the acquisitions of information systems.

ROI is a performance measure used to evaluate the efficiency of investment or to compare the efficiency of a number of different investments. To calculate ROI, the net benefit (return) of an investment is divided by the cost of the investment; the result is expressed as a percentage or ratio (Erdogmus, Favaro and Strigel 2004)

There are many other ROI definitions in the literature (e.g. (Return on Investment (ROI), Glossary n.d.; Mogollon & Raisinghani 2003)). Each definition focuses on certain ROI aspects. With all the diversity of the definitions, the primary notion is the same: ROI is a fraction, the numerator of which is "net gain" (return, profit, benefit) earned as a result of the project (activity, system operations), while the denominator is the "cost" (investment) spent to achieve the result.

In general, predicting future is notoriously prone to uncertainties and errors. Estimating future project costs and returns also is a challenging endeavor (Stamelos & Angelis 2001; Daneva & Wieringa 2008; Eckartz 2009; Jorgensen & Shepperd 2007). Due to a variety of reasons actual numbers usually differ from the ones estimated in advance. The errors in estimating costs and returns will propagate through the ROI formula and result in inaccuracies of the ROI evaluations.

Estimating the accuracy of the ROI evaluations should be considered an essential part of the ROI calculations because ROI is used to make critical business decisions. Neglecting to estimate ROI accuracy may lead to wrong decisions on acquisition of information systems.

The purpose of this study is to estimate the accuracy of the ROI evaluations. The study provides estimates of the ROI accuracy in the context of the information systems implementations. Although the focus of the research is on the information systems, significant part of it can be applied to other types of systems and other fields of ROI evaluations.

The research is intended to answer the following questions:

- How inaccuracies of determining project costs and benefits propagate through the ROI calculations and affect ROI accuracy?
- What levels of the quantitative error estimates of the ROI evaluations can be expected for typical scenarios of the information system implementations?

Several methodologies have been used to achieve the research objectives. Literature review method was used to gather and analyze information related to the accuracy of estimating project costs and returns. Measurements theory and error analysis, specifically, propagation of uncertainties methods were used to derive analytical expressions for ROI errors. Monte Carlo simulation methodology was used to design and deliver a quantitative experiment to model costs and returns estimating errors and calculate ROI accuracies. Spreadsheet simulation (Microsoft Excel spreadsheets enhanced with Visual Basic for Applications) was used to implement Monte Carlo simulations.

This research has the following scope and assumptions.

1. Most common definition treats ROI as a measure / metric / ratio / number (Erdogmus, Favaro and Strigel 2004). In some cases, return on investment is understood as a "method" or "approach" – "ROI analysis" (Mogollon & Raisinghani 2003; Andolsen 2004). This research is focused on the ROI as an individual measure.

2. ROI analysis can be performed with different purposes. As it was mentioned, ROI can provide rational for the future investments and acquisition decisions (e.g. project prioritization/ justification and facilitating informed choices about which projects to pursue). Evaluating future investments and making decisions on the information systems acquisitions are the processes based on the predicted data. By definition predicted data is likely to have certain level of variance from the amounts that will be really experienced later.

   To avoid unnecessary complications and focus on the ROI accuracy, it has been assumed that projects are relatively short-time efforts and value of money is not explicitly considered. Also, such effects as "negative benefits" (Lim et al 2011) or decrease of productivity immediately after implementation of a new information system are not considered.

3. Software effort/costs and benefits estimation methods are out of the research scope. It is assumed that appropriate methods were used to estimate costs and benefits, and the results are available to the ROI estimators.

4. The focus of the study is on the ROI accuracy. Higher –level aspects of ROI research, e.g. its positioning in the business value of information technology (IT) and information systems (IS), IS/IT valuation or benefit valuation/management – are out of the scope.

5. Other typical performance measures such as the net present value of IS/IT projects are out of the scope.

The results of this study are intended for researchers in information systems, technology solutions and business management, and also for information specialists, project managers, program managers, technology directors, and information systems evaluators. Most results are applicable to ROI evaluations in a wider subject area.

The importance of the problem is due to a wide use of the ROI evaluations in making investment decisions. The main contribution of the study is that this is the first systematic effort to evaluate ROI accuracy. Analytical expressions have been derived for estimating errors of the ROI evaluations. Results of the Monte Carlo simulation will help practitioners in making informed decisions based on explicitly stated factors influencing the ROI uncertainties. Also, the paper contributes to more accurate ROI evaluations by drawing evaluators' attention to the ways of minimizing evaluation errors.

The paper is structured as follows. Section 1 provides a brief introduction, outlines research objectives, defines methodology, and identifies limitations and assumptions of the study. Section 2 reviews previous work on ROI. Section 3 analyzes how uncertainties propagate through the ROI formula. The author derives mathematical approximations for the ROI accuracy by applying accepted approaches from measurements theory. In Section 4, the author applies a Monte Carlo simulation to illustrate the main implications of the study. The evidence is presented that the errors for ROI estimates are considerably high and that they should be taken account when making IT decisions. Analytic and simulation results discussed in Section 5. The paper concludes with final remarks in Sections 6.

## 2.0 Literature Review

A literature review has been conducted in support of this research. The review didn't reveal any papers specifically investigating methods of estimating ROI accuracy or case studies on this topic.

Two articles deal with the ROI accuracy (Botchkarev & Andru 2011; Andru & Botchkarev 2011). The value of these articles is in demonstrating the approach, and illustrating the level of the ROI accuracy for a typical CRM project. Accuracy assessment of the ROI calculations was performed on a specific example. Though not claiming any generic value, it was shown that even relatively low-level errors of estimating costs and returns (+/- 10%) may lead to significant ROI inaccuracies. That led to a conclusion that to make ROI number meaningful, it should be provided with an assessment of its accuracy.

Further literature review was focused on the accuracy of the components used to calculate ROI: costs and financial returns/benefits (Botchkarev 2015). The review indicates that in 75% of the projects the cost error estimates fall within the range of 20% to 60% with most likely value of error from 30% to 50%. The literature review didn't reveal any studies neither on the methodology of estimating accuracy of predicted benefits nor on actual numbers based on the case studies. The assumption was drawn that the same (or larger) quantitative levels of benefits estimation accuracy could be expected as we experience for cost estimation accuracy.

## 3.0 Analytical Estimation of the ROI Accuracy

The ROI is defined as:

$$R_{est} = \frac{B_{est} - C_{est}}{C_{est}} \qquad (1)$$

where $C_{est}$ is an estimate of the cost to implement a project (predicted cost);

$B_{est}$ is an estimate of the benefit (financial return) from the project implementation (predicted benefit);

$R_{est}$ is the value of the ROI calculated based on the estimated costs and benefits (predicted ROI).

Equation (1) represents a complex non-linear function. Due to the uncertainties of the estimation process, actual costs ($C_{act}$) and actual benefits ($B_{act}$), realized after the project is completed, will be different from the estimated ones. Because of multiple impacting uncertainties the absolute estimating errors could be considered random and expressed as follows:

$$\delta C = C_{act} - C_{est}; \quad \delta B = B_{act} - B_{est}$$

Hence, the actual ROI will also be different from the estimated one. The error of estimating ROI can be written as:

$$\delta R = R_{act} - R_{est}$$

The problem is to define an analytical expression for the ROI estimation error as a function of the uncertainties measuring costs and benefits:

$$\delta R = f(\delta C, \delta B)$$

or for the relative ROI error:

$$\frac{\delta R}{R_{act}} = F\left(\frac{\delta C}{C_{act}}, \frac{\delta B}{B_{act}}\right)$$

Similar problem is well-known in the physical sciences and engineering, and studied in the measurements theory and error analysis (Taylor 1997; Hughes & Hase 2010). In measurements, involving readings from two or more physical devices/meters, there is a need to assess the error of the experimental result when the readings are combined in an equation, e.g. three sides of a block are measured with a tape measure and then the volume of the block is calculated by multiplying these readings and the volume of the block is determined. Uncertainties that occurred in measuring the sides will propagate through the equation/formulae and affect the uncertainty of the calculated result. Usually, this area of studies is called error propagation or propagation of uncertainties and it is based on the mathematics of stochastic processes and, specifically, on algebra of stochastic variables. Measurement theory developed certain methods of calculating output errors depending on the type of the equations/formulae used: whether the measured parameters are added, deducted, multiplied, etc. This research follows the considerations accepted in the measurements theory. However, it should be noted that some assumptions and subsequent mathematical approximations common for the measurement field (e.g. the absolute error of the measurement is much smaller than the value of the measured quantity) may not be valid for all ROI evaluation scenarios. So, error analysis mathematics should be applied with caution.

**Maximum probable error – worst-case scenario.** Let's determine the maximum probable error for ROI. Maximum probable error represents a worst-case scenario: the errors assume largest possible values and in the most "undesirable" way, i.e. benefits are overestimated and costs are underestimated, or vice versa. Equation (1) can be rewritten to show maximum and minimum levels of the ROI

Maximum
$$R_{est} + \delta R = \frac{B_{est} + \delta B}{C_{est} - \delta C} - 1 \qquad (2)$$

Minimum
$$R_{est} - \delta R = \frac{B_{est} - \delta B}{C_{est} + \delta C} - 1 \qquad (3)$$

Equations (2) and (3) can be rearranged to find maximum probable error $\delta R$:

$$\delta R \approx \frac{B_{est}}{C_{est}} \left( \frac{\delta B}{B_{est}} + \frac{\delta C}{C_{est}} \right) \qquad (4)$$

Appendix A shows mathematical details of deriving (4). ROI maximum probable error approximately equals benefits-costs ratio multiplied by the sum of benefits and costs relative errors.

**Probable error.** Maximum probable error, presented in a previous subsection, dealt with a worst-case scenario. Although important and conceivable, this scenario will not occur often. In a more likely scenario, when errors are random and independent, errors of estimating benefits and costs will have different signs and may be partially compensating each other. This scenario also needs to be assessed.

A generalized formula for a probable error for a two-variable function R has been derived in (Taylor 1997 pp. 62, 141; Hughes & Hase 2010):

$$\delta R \approx \sqrt{\left(\frac{\partial R}{\partial B}\delta B\right)^2 + \left(\frac{\partial R}{\partial C}\delta C\right)^2} \qquad (5)$$

Substituting equation (1) into (5) and taking partial derivatives of the ROI function with respect of B and C, equation (5) can be transformed to

$$\delta R \approx \frac{B_{est}}{C_{est}}\sqrt{\left(\frac{\delta B}{B_{est}}\right)^2 + \left(\frac{\delta C}{C_{est}}\right)^2} \qquad (6)$$

Appendix B shows mathematical details of deriving (6). ROI probable error approximately equals benefits-costs ratio multiplied by the square root of the sum of squared benefits and costs relative errors.

**Breakdown of benefits and costs.** So far in this section to simplify the layout of the mathematical formulae, it was assumed that the value of the benefits (financial returns) is given by a single number $B_{est}$. For example, the project has a single type of benefits: cost savings due to downsizing, e.g. salaries and wages of the full time employees saved due to the system implementation. In practice, there could be a variety of the benefits types: e.g. increased revenues due to increased sales, or sales margins; revenue enhancement, e.g. additional revenues were gained due to better targeted marketed and advertising; revenue protection, e.g. imminent fine was avoided (due to demonstrated compliance with regulatory requirements). The same refers to the costs. Common cost types include: cost of software development or customization/configuration, cost of IT infrastructure, e.g. software/licenses - initial and annual maintenance; hardware - if IS run in-house (e.g. purchasing and installation of new servers); hosting - if information system provided as Software as a Service by a third party, cost of labour, etc.

So for a generic project, benefits $B_{est}$ and costs $C_{est}$ will be represented by summations of individual benefits and costs

$$B_{est} = \sum_i B_i; \qquad C_{est} = \sum_j C_j$$

where $B_i$ - $i$-th component of the financial return; and $C_j$ - $j$-th component of the system cost.

Most likely, each of these benefits and costs types will be estimated separately using different tools/methods, and have their own (specific) estimation error values, i.e. $\delta B_i$ and $\delta C_j$. As it is derived in (Taylor 1997; Hughes & Hase 2010), uncertainty propagation for the operation of summation can be estimated using the following formulae:

Maximum probable error
$$\delta B \approx \sum_i \delta B_i \qquad \delta C \approx \sum_j \delta C_j \qquad (7)$$

Probable error (sum in quadrature)
$$\delta B \approx \sqrt{\sum_i (\delta B_i)^2} \qquad \delta C \approx \sqrt{\sum_j (\delta C_j)^2} \qquad (8)$$

General procedure for estimating ROI errors starts with calculating overall errors of benefits and costs using equations (7) or (8) and then proceeds with substituting the results in equations (4) or (6).

## 4.0 Estimating ROI Accuracy with Monte Carlo Simulation

Monte Carlo simulation offers itself as a flexible technique for estimating ROI accuracy. It provides much more comprehensive insights into dependences of the costs and benefits uncertainties and ROI errors. Spreadsheet software packages have been widely used for Monte Carlo simulations due to their availability and simplicity (Chew and Walczyk 2012, Farrance and Frenkel 2014). In this study, the simulation was implemented on Microsoft Excel 2010 spreadsheets using Visual Basic for Applications (VBA). Earlier versions (1998, 2000, 2003 and 2007) of Excel were strongly criticized by the statistical community for their accuracy flaws (McCullough and Wilson 2005, McCullough and Heiser 2008). Recent research provides evidence that Excel 2010 demonstrates certain improvements, although still not perfect (Keeling and Pavur 2011, Mélard 2014, Kallner 2015). Known Excel limitations (specifically, relatively short cycle length and low numerical accuracy of certain statistical functions) are not critical for this application. The number of simulation trials and generated random numbers in the study is significantly smaller than the Excel cycle length – 2^24 (over 16 million). Also, there are no very small numbers or numbers that would differ in the fifth or sixth decimal place – issues that make Excel unsuitable in certain physical or mathematical sciences (Farrance and Frenkel 2014).

The Monte Carlo simulation process flowchart used in the study is shown in Figure 3. As a first step of setting a new case, a project cost value (used as an actual cost) was randomly selected from one of the three project ranges: small (100K-500K), medium (501K-900K) or large (901K-1,300K). Using the cost value, benefit amount was calculated at a certain benefit-cost ratio. Actual ROI was calculated using a standard formula:

$$R_{act} = \frac{B_{act} - C_{act}}{C_{act}}$$

Estimated ROI will differ from the actual value due to the uncertainties in estimating benefits and costs. These uncertainties were generated through a range of relative errors of benefits and costs $\delta B/B_{act}, \delta C/C_{act}$.

Upper and lower levels of the estimated benefits were calculated as follows

$$B_{estU} = B_{act} + B_{act}(\delta B/B_{act})$$

$$B_{estL} = B_{act} - B_{act}(\delta B/B_{act})$$

Then, estimated value of benefits $\beta_i$ was generated as a random number within the lower and upper bounds $\beta_i \in [B_{estL}, B_{estU}]$. Microsoft Excel VBA RND function was used to generate random numbers uniformly distributed within the specified interval.

Estimates of costs $\zeta_i$ were generated using the same approach $\zeta_i \in [C_{estL}, C_{estU}]$.

Estimated ROI values were calculated as

$$R_{esti} = \frac{\beta_i - \zeta_i}{\zeta_i}$$

Finally, ROI error $\delta R$ (mean absolute error), after N Monte Carlo iterations, was calculated as

$$\delta R = \frac{1}{N} \sum_{i}^{N} |R_{act} - R_{esti}|$$

Several cases were run to determine the required number of iterations (similar to the approach of Farrance and Frenkel 2014). The results demonstrated that the amount of the ROI error converges to the first or second decimal of a percent when the number of iterations reaches 15,000 to 20,000. As the runtime was not an issue (under 10 sec for a single point) due to a relatively simple model, the number of iterations was set to 30,000.

Results of the simulation are shown on the Figures 4-5. Fig. 4 shows dependences of the ROI error $\delta R$ with the increase of the relative errors of benefits and costs estimates $\delta B/B_{act}$, $\delta C/C_{act}$ for the errors in the range from 0 to 45%. Fig. 5 shows similar data for the larger errors: 40% to 95%.

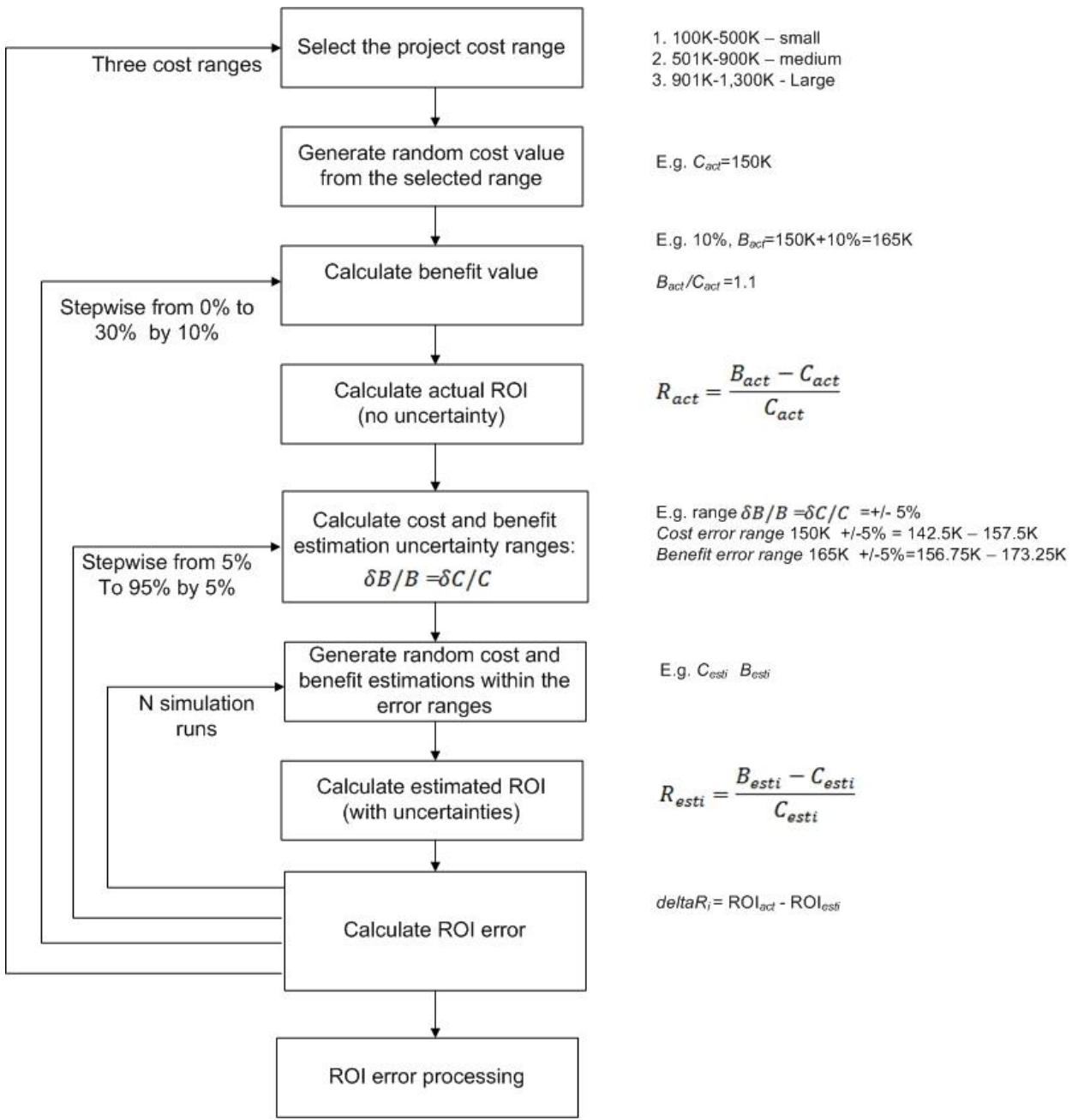

*Figure 3.* Simulation Process Flowchart

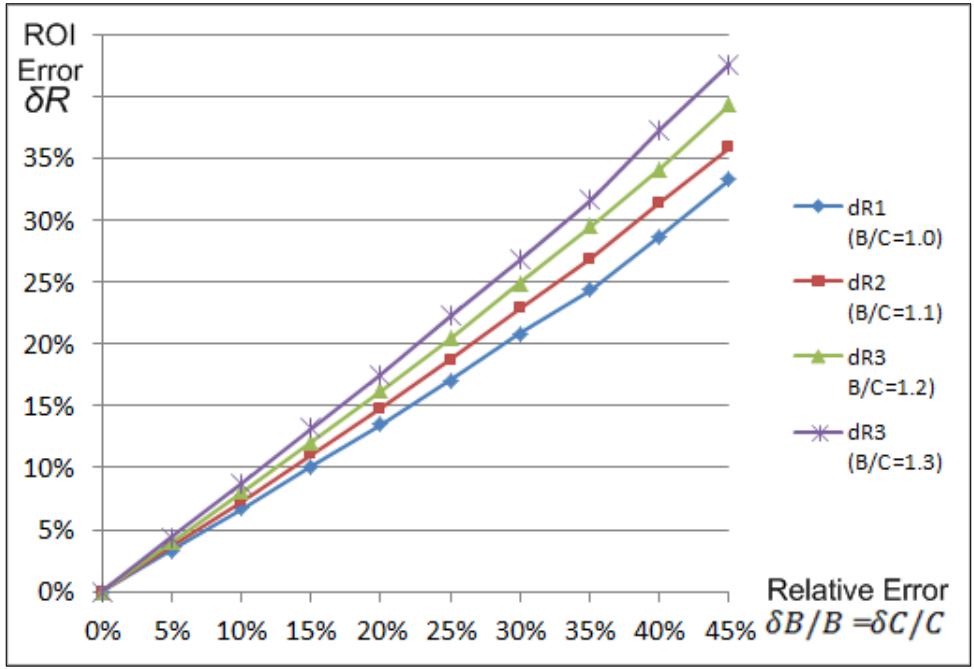

*Figure 4.* ROI error for the lower-level benefits and costs relative errors

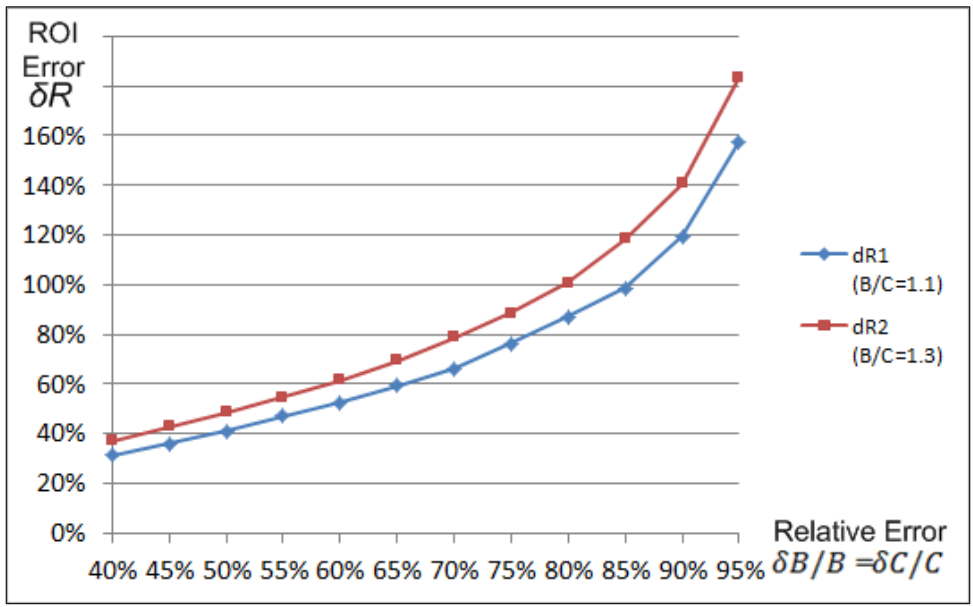

*Figure 5.* ROI error for the higher-level benefits and costs relative errors

## 5.0 Discussion

Analytical expressions for the ROI errors derived in Section 3.0 are based on certain assumptions and simplifications. The prime one is that benefits and costs estimating errors are small and Taylor series expansion can be used. It is demonstrated in Appendix A, Fig. A.1, that variance between the exact and approximated solutions increases rapidly when value of the relative error exceeds 15-20%. This data suggests that approximated expressions for the ROI errors are best used for relative errors under 15-20%. It should be noted that the approximated line goes below the exact line. As a result, approximated errors may underestimate real ROI errors.

Analytical expressions (with better ROI accuracy) for the cases with larger errors of costs and benefits are difficult to derive. There are studies in this area, e.g. (Seiler 1987), but the complexity of the solutions precludes them from being recommended to practitioners.

Results of the simulation presented in the Section 4 show how the ROI absolute mean error is changing with the relative errors of benefits and costs. The behaviour of the graphs is different for the lower and higher levels of the relative errors. For better visual perception they are demonstrated separately. The graph for the lower error levels (see Fig. 5) shows almost linear relationship between the ROI absolute error and relative errors of benefits and costs (especially when relative errors are under 30%). The graph for the higher error levels (over 40%) shows exponential growth (see Fig. 5). As it might be expected, simulation has shown no difference for the ROI error behaviour for the projects of different sizes. The results show that ROI errors for the small and large projects (for the same relative errors of benefits and costs) are identical.

Simulation results include the assumption that the relative errors of benefits and costs are equal (to ensure better visual presentation). Also, the distribution of the relative errors of benefits and costs was set to be uniform.

To round off the Discussion section, it is important to note that as any project is a unique endeavour (by definition), the same characteristic applies to the value of ROI errors in each project. It means that there are no any standard or expected ROI error amounts. Everything depends on how accurate were the financial assessments of the project benefits and costs. Project manager or analyst has to make ROI error estimations in specific conditions of the project. The results of this study provide a foundation for such estimations.

When assessing the ROI uncertainty, it is also noteworthy to take into account the ultimate financial implications not the intermediate parameters. For example, a company is developing a new software solution. The workload has been estimated with uncertainty of +/-50%. It seems at this point that expected ROI error will also be very large. And it is true, if the project would be developed in-house and workload will be directly translated into costs with the similar errors. However, if the software development would be outsourced through a fixed-price contract – the financial/cost uncertainty for the company will be close to zero, and so will be ROI error.

## 6.0 Concluding Remarks

Estimating accuracy of the ROI evaluations should become a part of the ROI assessments' best practices in order to avoid erroneous investment decisions. This study provided the first (to the best knowledge of the author) systematic research (both analytical and using simulation) of the accuracy of the ROI evaluations in the context of the information systems implementations and laid foundation for further theoretical and practical works in this area.

Future research may be focused on developing a framework of assessing and presenting benefits accuracy in a more standardized way. Also, research can be conducted into mathematical aspects of estimating ROI accuracy in the cases when estimating errors of benefits and costs are large, and have various probability distribution functions.

**Appendix A. Analytical Derivation of Maximum Probable Error**

In the equation (1), a variable $(B_{est})$ is used more than once. That may lead to an effect of errors cancelling themselves (i.e. compensating errors) (Taylor 1997, p. 74). We can re-arrange equation (1) to avoid using a variable more than once

$$R_{est} = \frac{B_{est}}{C_{est}} - 1 \tag{A.1}$$

According to (Taylor 1997 p. 66), any problem for propagation error can be subdivided into sequence of steps, each of them based on the elementary mathematical operation. The second term in equation (A.1) does not include error component and could be neglected in the further error analysis. The first term is a quotient of two variables and error propagation for such a function is well-known (Taylor 1997; Hughes & Hase 2010). The maximum value of the ROI in equation (A.1) will occur when the numerator will be maximum and denominator will be minimum:

$$R_{est} + \delta R = \frac{B_{est} + \delta B}{C_{est} - \delta C} \tag{A.2}$$

Minimum value can be expressed as

$$R_{est} - \delta R = \frac{B_{est} - \delta B}{C_{est} + \delta C} \tag{A.3}$$

Following (Lindberg 2000; Physics Laboratory Companion), we can rewrite equation (A.2)

$$B_{est} + \delta B = (R_{est} + \delta R)(C_{est} - \delta C) =$$

$$R_{est}C_{est} - R_{est}\delta C + C_{est}\delta R - \delta R \delta C$$

Assuming the errors are small, the last term ($\delta R \delta C$) can be neglected, and absolute ROI error can be written as

$$\delta R \approx (B_{est} + \delta B - R_{est}C_{est} + R_{est}\delta C)/C_{est} \tag{A.4}$$

Taking into account that $R_{est} = B_{est}/C_{est}$ and substituting into equation (A.4), the expression for the maximum probable absolute error will be:

$$\delta R \approx \frac{C_{est}\delta B + B_{est}\delta C}{C_{est}^2} \tag{A.5}$$

or, multiplying both numerator and denominator by $B_{est}$, and rearranging

$$\delta R \approx \frac{B_{est}}{C_{est}}\left(\frac{\delta B}{B_{est}} + \frac{\delta C}{C_{est}}\right) \tag{A.6}$$

As it is observed in (Taylor 1997; Hughes & Hase 2010; Physics Laboratory Companion), error for a quotient is better expressed in terms of the relative error. Dividing both parts of equation (A.6) by $R_{est}$, we get the following formula

$$\frac{\delta R}{|R_{est}|} \approx \frac{\delta B}{B_{est}} + \frac{\delta C}{C_{est}} \tag{A.7}$$

We arrived at a formula that is commonly used in the error propagation assessments for quotients (Taylor 1997; Hughes & Hase 2010; Physics Laboratory Companion).

Another approach to calculate maximum probable error is as follows. Equation (A.1) may be rewritten to show maximum and minimum levels of the ROI

Maximum
$$R_{est} + \delta R = \frac{B_{est} + \delta B}{C_{est} - \delta C} - 1 \tag{A.8}$$

Minimum
$$R_{est} - \delta R = \frac{B_{est} - \delta B}{C_{est} + \delta C} - 1 \tag{A.9}$$

Following a method used in (Taylor 1997 pp. 51; Palmer n.d.), equation (A.9) can be rewritten as

$$R_{est} + \delta R = \frac{B_{est}}{C_{est}}\left(\frac{1 + \delta B/B_{est}}{1 - \delta C/C_{est}}\right) - 1 \tag{A.10}$$

Assuming the errors are small and using a binomial theorem, a component of (A.10) can be simplified (approximated by a Taylor series)

$$\frac{1}{1 - \delta C/C_{est}} \approx 1 + \delta C/C_{est} + (\delta C/C_{est})^2 + \cdots$$

Using only the first two terms of the approximation, equation (A.10) can be rewritten as

$$R_{est} + \delta R \approx \frac{B_{est}}{C_{est}}\left(1 + \frac{\delta B}{B_{est}}\right)\left(1 + \frac{\delta C}{C_{est}}\right) - 1 \qquad (A.11)$$

Rearranging equation (A.11), the error can be expressed as

$$\delta R \approx \frac{B_{est}}{C_{est}}\left(1 + \frac{\delta B}{B_{est}} + \frac{\delta C}{C_{est}} + \frac{\delta B}{B_{est}}\frac{\delta C}{C_{est}}\right) - 1 - R_{est}$$

Assuming again that the relative errors are small, so the last term in the brackets can be neglected and substituting $R_{est} = (B_{est}/C_{est}) - 1$

$$\delta R \approx \frac{B_{est}}{C_{est}}\left(1 + \frac{\delta B}{B_{est}} + \frac{\delta C}{C_{est}}\right) - 1 - \frac{B_{est}}{C_{est}} + 1 =$$

(A.12)

$$\frac{B_{est}}{C_{est}}\left(\frac{\delta B}{B_{est}} + \frac{\delta C}{C_{est}}\right)$$

Similar results can be gained if we use a generalized formula for a maximum probable error which for our case could be expressed through the total differential of a function (Taylor 1997 pp. 51; Palmer n.d.)

$$dR = \left(\frac{\partial R}{\partial B}\right)dB + \left(\frac{\partial R}{\partial C}\right)dC$$

Assuming $dR = \delta R$, and likewise for the other differentials, and that the variables $C$ and $B$ are independent, the result for errors

$$\delta R \approx \left|\frac{\partial R}{\partial B}\right|\delta B + \left|\frac{\partial R}{\partial C}\right|\delta C \qquad (A.13)$$

Formula (A.13) neglects higher order derivatives of the function which is considered a good approximation when the errors are small.

Substituting equation (A.1) into (A.13) and taking partial derivatives of the ROI function with respect of B and C

$$\delta R \approx \left|\frac{\partial}{\partial B}\left(\frac{B_{est}}{C_{est}} - 1\right)\right| \delta B + \left|\frac{\partial}{\partial C}\left(\frac{B_{est}}{C_{est}} - 1\right)\right| \delta C =$$

$$\left|\left(\frac{1}{C_{est}}\delta B\right)\right| + \left|B_{est}\frac{\partial}{\partial C}\left(\frac{1}{C_{est}}\right)\right| \delta C =$$

$$\left|\frac{1}{C_{est}}\delta B\right| + \left|B_{est}\left(-\frac{1}{C_{est}^2}\delta C\right)\right| = \qquad (A.14)$$

$$\frac{C_{est}\delta B + B_{est}\delta C}{C_{est}^2} =$$

$$\frac{B_{est}}{C_{est}}\left(\frac{\delta B}{B_{est}} + \frac{\delta C}{C_{est}}\right)$$

We can observe that equations (A.6), (A.12) and (A.14) provide the same result.

It should be noted that equation (A.12) for the ROI maximum probable error was derived using the first two items in the Taylor series expansion:

$$\frac{1}{1 - \delta C/C_{est}} \approx 1 + \delta C/C_{est} \qquad (A.15)$$

Fig A.1 demonstrates the graphs for the left (exact) and right (approximated) parts of the equation (A.15) for a range of the cost relative errors $\delta C/C_{est}$. M is a numeric value of the approximated term.

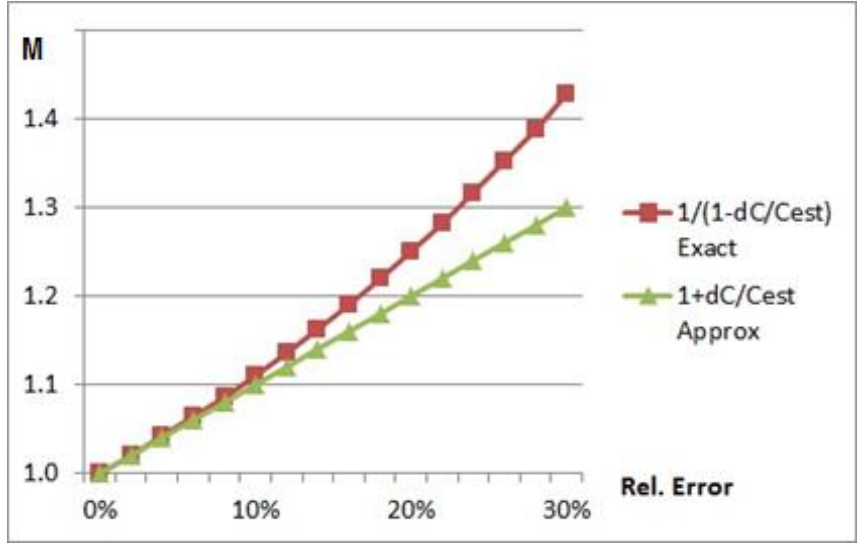

*Figure A.1.* Taylor series expansion

The graph shows that variance between the exact and approximated solutions increases rapidly when value of the relative error exceeds 15-20%.

**Appendix B. Analytical Derivation of Probable Error**

Substituting equation (1) into (5) and taking partial derivatives of the ROI function with respect of B and C, equation (5) can be transformed to

$$\delta R \approx \sqrt{\left(\frac{\partial R}{\partial B}\delta B\right)^2 + \left(\frac{\partial R}{\partial C}\delta C\right)^2} =$$

$$\sqrt{\left[\frac{\partial}{\partial B}\left(\frac{B_{est}}{C_{est}} - 1\right)\delta B\right]^2 + \left[\frac{\partial}{\partial C}\left(\frac{B_{est}}{C_{est}} - 1\right)\delta C\right]^2} =$$

$$\sqrt{\left(\frac{1}{C_{est}}\delta B\right)^2 + \left[B_{est}\frac{\partial}{\partial C}\left(\frac{1}{C_{est}}\right)\delta C\right]^2} =$$

$$\sqrt{\left(\frac{1}{C_{est}}\delta B\right)^2 + \left[B_{est}\left(-\frac{1}{C_{est}^2}\delta C\right)\right]^2} =$$

$$\sqrt{\frac{C_{est}^2 \delta B^2 + B_{est}^2 \delta C^2}{C_{est}^4} * \frac{B_{est}^2}{B_{est}^2}} =$$

$$\frac{B_{est}}{C_{est}}\sqrt{\left(\frac{\delta B}{B_{est}}\right)^2 + \left(\frac{\delta C}{C_{est}}\right)^2}$$

**Reference List**

Andolsen, A.A. (2004). Investing Wisely for the Future, *The Information Management Journal*, *8*(5), 47-54.

Andru, P., Botchkarev, A. (2011). The Use of Return on Investment ROI) in the Performance Measurement and Evaluation of Information Systems. *Information Management, Access and Privacy Symposium.* Presentation available at E-prints in Library and Information Science E-LIS), http://hdl.handle.net/10760/15503

Botchkarev, A., & Andru, P. (2011). A return on investment as a metric for evaluating information systems: Taxonomy and application. *Interdisciplinary Journal of Information, Knowledge, and Management, 6*, 245-269. http://www.ijikm.org/Volume6/IJIKMv6p245-269Botchkarev566.pdf

Botchkarev, A. (2015). Accuracy of estimating project costs and benefits: an overview of research in information systems. *Journal of Emerging Trends in Computing and Information Sciences*. (submitted)

Chew, G., & Walczyk, T. (2012). A Monte Carlo approach for estimating measurement uncertainty using standard spreadsheet software. *Analytical and bioanalytical chemistry*, *402*(7), 2463-2469.

Daneva, M., and Wieringa, R. (2008). "Cost estimation for cross-organizational ERP projects: research perspectives" *Software Quality Journal, 16*(3), 459-481. doi:10.1007/s11219-008-9045-8

Eckartz, S. M. (2009). Costs, Benefits and Value Distribution - Ingredients for Successful Cross-Organizational ES Business Cases. In: *32nd Information Systems Research Seminar in Scandinavia, IRIS 32, Inclusive Design*, 9-12 August 2009, Molde, Norway.

Erdogmus, H., Favaro, J. and Strigel, W. (2004). Guest Editors' Introduction: Return on Investment. *IEEE Software, 21*(3), 18-22. doi:10.1109/MS.2004.1293068
http://www.computer.org/csdl/mags/so/2004/03/s3018.html

Farrance, I., & Frenkel, R. (2014). Uncertainty in Measurement: A Review of Monte Carlo Simulation Using Microsoft Excel for the Calculation of Uncertainties Through Functional Relationships, Including Uncertainties in Empirically Derived Constants. *The Clinical Biochemist Reviews*, *35*(1), 37.
http://www.ncbi.nlm.nih.gov/pmc/articles/PMC3961998/pdf/cbr-35-37.pdf

Hughes, I., & Hase, T. (2010). Measurements and their uncertainties: a practical guide to modern error analysis. Oxford University Press.

Jorgensen, M., Shepperd, M. (2007). A Systematic Review of Software Development Cost Estimation Studies. *IEEE Transactions on Software Engineering*, *33*(1), 33 -53. doi:10.1109/TSE.2007.256943

Kallner, A. (2015). Microsoft EXCEL 2010 offers an improved random number generator allowing efficient simulation in chemical laboratory studies. *Clinica chimica acta; international journal of clinical chemistry*, *438*, 210-211.

Keeling, K.B and Pavur, R.J. (2011). Statistical Accuracy of Spreadsheet Software, *The American Statistician, 65*(4), 265-273, http://dx.doi.org/10.1198/tas.2011.09076
http://www.tandfonline.com/doi/pdf/10.1198/tas.2011.09076

Lim, J. Y., Kim, M. J., Park, C. G., & Kim, J. Y. (2011). Comparison of Benefit Estimation Models in Cost-Benefit Analysis: A Case of Chronic Hypertension Management Programs. *Journal of Korean Academy of Nursing*, *41*(6), 750-757.

http://synapse.koreamed.org/search.php?where=aview&id=10.4040/jkan.2011.41.6.750&code=0006JKAN&vmode=FULL

McCullough, B. D., & Heiser, D. A. (2008). On the accuracy of statistical procedures in Microsoft Excel 2007. *Computational Statistics & Data Analysis*, *52*(10), 4570-4578.

McCullough, B. D., & Wilson, B. (2005). On the accuracy of statistical procedures in Microsoft Excel 2003. *Computational Statistics & Data Analysis*, *49*(4), 1244-1252.

Mélard, G. (2014). On the accuracy of statistical procedures in Microsoft Excel 2010. *Computational Statistics*, 1-34.

Mogollon, M. and Raisinghani, M. (2003). Measuring ROI in E-Business: A Practical Approach. *Information Systems Management*, *20*(2), 63 - 81.

Return On Investment (ROI), Glossary. Centers for Disease Control and Prevention. http://www.cdc.gov/leanworks/resources/glossary.html accessed June 27, 2014

Seiler, F. A. (1987). Error propagation for large errors. *Risk Analysis, 7*(4), 509-518.

Stamelos, I., Angelis L. (2001). Managing uncertainty in project portfolio cost estimation. *Information and Software Technology*. *43*(13), 759-768. doi:10.1016/S0950-5849(01)00183-5

Taylor, J. R. (1997). *An introduction to error analysis: the study of uncertainties in physical measurements*. University science books. 2nd edition.